\documentclass[twocolumn,prl,superscriptaddress]{revtex4-2}
\usepackage{float}
\usepackage{graphicx}
\usepackage{amssymb,amsmath}
\usepackage{bm}
\usepackage{dcolumn}

\usepackage{hyperref}
\hypersetup{backref,colorlinks=true,breaklinks,urlcolor=blue,linkcolor=blue,citecolor=blue}
\usepackage{xcolor}

\usepackage[capitalise]{cleveref}
\usepackage[notrig]{physics}   
\usepackage{nicefrac}
\usepackage{tikz}

\newcommand{\eff}{\mathrm{eff}}
\renewcommand{\i}{\mathrm i}

\makeatletter
\def\maketitle{
	\@author@finish
	\title@column\titleblock@produce
	\suppressfloats[t]}
\makeatother

\begin{document}

\title{Quantum trajectories for time-local non-Lindblad master equations}
\author{Tobias Becker}
\email{tobias.becker@tu-berlin.de}
\affiliation{Institut f\"ur Theoretische Physik, Technische Universit\"at Berlin, Hardenbergstrasse 36, 10623 Berlin, Germany}
\author{Ch\'{e} Netzer}
\affiliation{Institut f\"ur Theoretische Physik, Technische Universit\"at Berlin, Hardenbergstrasse 36, 10623 Berlin, Germany}
\author{Andr\'{e} Eckardt}
\email{eckardt@tu-berlin.de}
\affiliation{Institut f\"ur Theoretische Physik, Technische Universit\"at Berlin, Hardenbergstrasse 36, 10623 Berlin, Germany}

\begin{abstract}
For the efficient simulation of open quantum systems we often use quantum jump trajectories given by pure states that evolve stochastically to unravel the dynamics of the underlying master equation. In the Markovian regime, when the dynamics is described by a Gorini-Kossakowski-Sudarshan-Lindblad (GKSL) master equation, this procedure is known as Monte-Carlo wavefunction (MCWF) approach . However, beyond ultraweak system-bath coupling, the dynamics of the system is not described by an equation of GKSL type, but rather by the Redfield equation, which can be brought into pseudo-Lindblad form. Here negative dissipation strengths prohibit the conventional approach. To overcome this problem, we propose a pseudo-Lindblad quantum trajectory (PLQT) unraveling. It does not require an effective extension of the state space, like other approaches, except for the addition of a single classical bit. We test the PLQT for the eternal non-Markovian master equation for a single qubit and an interacting Fermi Hubbard chain coupled to a thermal bath and discuss its computational effort compared to solving the full master equation. 
\end{abstract}

\maketitle
\emph{Introduction.--} Away from thermodynamic equilibrium the properties of an open quantum system do not simply follow from the fundamental principles of statistical mechanics, but depend on the very details of the surrounding environment. 
This includes both transient dynamics, as the algorithm of a quantum computer or the relaxation following a quantum quench, and non-equilibrium steady states. Therefore, it is crucial to find an effective equation of motion for the open system that accurately captures the impact of the environment. At the same time, and equally importantly, the theoretical description should allow for efficient numerical simulations. A powerful approach for the latter are quantum trajectory simulations, where a stochastic process for the evolution of pure states is considered, the ensemble average of which describes the open system. Compared to the evolution of the full density operator (scaling quadratically with the state-space dimension $D$), these simulations require less memory, since pure states scale only linearly with $D$. Moreover, such unravelings can also directly describe stochastic processes of measured systems \cite{THBasche1995,SGleyzes2007,ZKMinev2019}. 

Quantum trajectory simulations are rather straightforward in the ultraweak-coupling limit, where the system-bath coupling is weak compared to the (quasi)energy level splitting in the system. In this case, the system is described by a master equation of GKSL (Gorini-Kossakowski-Sudarshan-Lindblad) form \cite{GoriniKossakowski1976,GLindblad1976} ($\hbar=1$),  
\begin{align}
	\dot \varrho = -\i[H, \varrho] + \sum\limits_{i} \gamma_i \left( L_i \varrho L_i^\dagger - \frac{1}{2} \{ L_i^\dagger L_i,\varrho \} \right) ,
	\label{eq:pseudo_lindblad}
\end{align}
with the coherent evolution captured by some Hamiltonian $H$ and dissipation described by jump operators $L_i$ with associated non-negative strengths $\gamma_i$. Here, $H$, $\gamma_i$, and $L_i$ can be time dependent. 

From this equation, we can immediately obtain a stochastic process for the evolution of pure states, known as Monte-Carlo wavefunction (MCWF) approach~\cite{Dalibard92,Kmolmer1993,GardinerZoller1992:1,GardinerZoller1992:2,NGisin1992:1,NGisin1992:2,HCarmicheal1999,ADaley2014}. In each time step $\delta t$, the state either evolves coherently according to  $\ket{\psi(t+\delta t)} \propto (1-\i \delta t H_\mathrm{eff}(t))\ket{\psi(t)}$  with probability $1- \sum_i r_i(t) \delta t$ and effective Hamiltonian 
\begin{align} 
	H_\mathrm{eff}(t) = H - \frac{\i}{2} \sum_i \gamma_i L_i^\dagger L_i,
	\label{eq:eff_Hamiltonian}
\end{align}
or a quantum jump occurs, $\ket{\psi(t+\delta t)} \propto L_i \ket{\psi(t)}$, with probability $r_i(t) \delta t$, with jump rates $r_i(t) = \gamma_i \matrixel{\psi(t)}{L_i^\dag L_i }{\psi(t)}$. The state of the system is then given (or approximated) by the ensemble average $\rho(t)= \overline{\dyad{\psi(t)}}$ over an infinitely (or sufficiently) large number $N$ of trajectories $\ket{\psi_n(t)}$, where $\overline{X} \equiv\frac{1}{N}\sum_{n=1}^N X_n$. 

However, the assumption of ultraweak coupling is questionable in various situations, for instance in large systems, with small finite-size gaps and tiny avoided crossings between many-body states, as well as in Floquet systems with driving frequency $\omega$, where the average quasi energy level spacing is $\omega/D$ \cite{eckardtColloquiumAtomicQuantum2017}. 

Beyond ultraweak coupling, master equations in pseudo-Lindblad form can be found, which look like a GKSL master equation \cref{eq:pseudo_lindblad}, except for the fact that the coefficients $\gamma_i$ also take negative values. For instance, the Redfield equation obtained in (Floquet)-Born-Markov approximation can be brought to this form \cite{TBecker2021}. Generally, negative relaxation strengths are relevant for non-Markovian dynamics \cite{DCruscinskiJPiiloWStrunz2017}, stochastic Hamiltonians with non-Markovian noise \cite{Groszkowski2022}, gauge transformed Lindbladians \cite{McDonaldClerk2023arxiv} and exact master equations \cite{KarrleinGrabert97,AShajiECGSudarshan2005,SAlipour2020}. These negative values are incompatible with the conventional MCWF, since the probability $r_i(t) \delta t$ for a quantum jump would become negative. To overcome this problem different quantum jump unravelings have been formulated, which, however, require significantly more computational resources \cite{breuerpetruccione1999,IKondovMschreiber2003,breuer2004,SuessEisfeldStrunz2014,MRHush2015,Piilo2009,KHaerkoenen_2010,KLuoma2020,SmirnePiilo2020}. In many approaches the system’s state space needs to be extended, so that its dimensionality at least doubles \cite{breuerpetruccione1999,IKondovMschreiber2003,breuer2004,ChinRivasHuelgaPlenio2010,SuessEisfeldStrunz2014,MRHush2015,StrasbergSchallerBrandes2016,PolettiSchaller2021}. For oscillating strengths between positive and negative values, moreover, an alternative non-Markovian quantum jump method (NMQJ) has been proposed in which jump processes are inverted \cite{PiiloManiscalco2008,Piilo2009,KHaerkoenen_2010,KLuoma2020}. This method does not work, if $\gamma_i<0$ for all times and does not admit independent (i.e.~parallel) evaluation of trajectories. For non-oscillatory strengths the rate operator quantum jump approach can be applied \cite{SmirnePiilo2020}, however, it requires a rather costly diagonalization of a state-dependent operator in every time step of the evolution. A generalization of the non-Markovian jump method to many-body systems has been proposed in Ref.~\cite{CGiulianoDPoletti2023} and can be used to study measurement induced phase transitions.

In this work we propose pseudo-Lindblad quantum trajectories (PLQT), which work for arbitrary $\gamma_i$, where the trajectories evolve independently and which does not require the doubling of state space. In the following, this is realized by extending the system’s state space in a minimal (and for the memory requirement of simulations practically irrelevant) fashion by a single classical bit $s\in\{-1,+1\}$, $\ket{\psi(t)} \to \{\ket{\psi(t)}, s(t)\}$.

\emph{Algorithm.--}
To unravel the dynamics of a pseudo-Lindblad quantum master equation by quantum trajectories $\{ \ket{\psi(t)} , s(t) \}$, first choose a time step $\delta t$, which is sufficiently short for the first-order time integration, and jump rates $r_i(t)>0$ for each jump operator $L_i$ (to be specified below). Within one time step a quantum jump occurs described by
\begin{align}
	\begin{aligned}
		\ket*{\psi^{(i)}(t+ \delta t)} &= \frac{\sqrt{|\gamma_i|} L_i \ket{\psi(t)}} {\sqrt{r_i(t)}} , \\
		s^{(i)}(t+ \delta t) &= \frac{\gamma_i}{|\gamma_i|}\, s(t),
		\label{eq:jump}
	\end{aligned}
\end{align} 
with probability $r_i(t)\delta t$
or, alternatively, with the remaining probability $1-\sum_i r_i(t)  \delta t$ the state evolves coherently, with $H_\mathrm{eff}$ [\cref{eq:eff_Hamiltonian}]  \cite{footnote_det}
\begin{align}
	\ket*{\psi^{(0)}(t+\delta t)} &= \frac{(1 -\i \delta t  H_\eff(t)) \ket{\psi(t)}}{\sqrt{1-\delta t \sum_i r_i(t)}} , 	\label{eq:stQTdet} \\
	s^{(0)}(t+\delta t)&=s(t).
\end{align} 

We now show that the ensemble of pure states obtained by this PLQT approach, converges to the correct density operator solving \cref{eq:pseudo_lindblad},
\begin{align}
	\varrho(t) = \overline{s(t) \dyad{\psi(t)}}.
\end{align}
For a pure initial state, $\sigma(t) = s(t) \dyad{\psi(t)}$, on average the update scheme is the weighted sum of the processes described above:
\begin{align}
	\begin{aligned}
		\overline{ \sigma (t+  \delta t)} =& \sum_i r_i(t) \delta t\  \sigma^{(i)}(t +\delta t) \\ &+ \bigg(1-\sum_i r_i(t) \delta t\bigg)  \sigma^{(0)}(t +\delta t),
	\end{aligned}
\end{align}
with $\sigma^{(i)} = s^{(i)} \dyad{\psi^{(i)}}$ and $\sigma^{(0)} = s^{(0)} \dyad{\psi^{(0)}}$. By inserting \cref{eq:jump,eq:stQTdet}, the jump rates $r_i(t)$ cancel out and one arrives at 
\begin{align}
	\begin{split}
		\overline{ \sigma (t+  \delta t)} =& \sigma(t) + \delta t \bigg( \sum_i \gamma_i L_i \sigma(t) L_i^\dagger \\ & - \i H_\mathrm{eff}(t) \sigma(t) + \i \sigma(t) H_\mathrm{eff}(t)^\dagger \bigg),
	\end{split}
	\label{eq:process_average}
\end{align}
almost corresponding to the action of the master \cref{eq:pseudo_lindblad}. 
The final step to arrive at \cref{eq:pseudo_lindblad} is to average \cref{eq:process_average} also over an ensemble of pure states at time $t$, so that $\overline{\sigma(t+\delta)}\to\varrho(t+\delta t)$ and $\sigma(t)\to\varrho(t)$.
As will be discussed below, one consequence of the presence of negative weights $\gamma_i<0$ is that individual wave functions $\psi_n$ are not normalized. As a result, the ensemble averaged trace is preserved only in the limit $N\to\infty$ of an infinite ensemble \footnote{The fluctuations of the trace and their dependence on $N$ are shown in Fig. 1 of the supplemental material \cite{supp_TBeckerCNetzerAEckardt2023} for an explicit example.}.
Therefore, in a finite ensemble, one obtains better convergence by explicit normalization, $\varrho_N = \frac{1}{\mathcal{N}}  \sum_n^N s_n \dyad{\psi_n}$, with $\mathcal{N} = \sum_n^N s_n \braket{\psi_n}{\psi_n}$ at every time $t$.
A rigorous proof of our algorithm using the Ito-formalism is outlined in the supplemental material \cite{supp_TBeckerCNetzerAEckardt2023}. In case that all $\gamma_i$ are positive, the sign bits do not change and the algorithm corresponds to the conventional MCWF approach. 

Note that, recently, also another unraveling of non-Lindblad master equations was proposed in Ref.~\cite{DonvilMGinanneschi2022}. It is different from our approach, but also involves an effective classical degree of freedom, given by a real number of constant average, rather than our single bit, whose average is time-dependent, as will be seen below. 

For the PLQT approach, as for other unraveling schemes \cite{IKondovMschreiber2003}, the jump rates $r_i(t) > 0$ can, in principle, be chosen arbitrarily. In practice there is, however, a trade off. Whereas for too small rates $r_i$, large ensembles of trajectories are required to sample each jump process $i$ sufficiently, we also have to require that the probability $1- \sum_i r_i\delta t$ remains positive and large enough for the given time step $\delta t$. A typical choice is \cite{ADaley2014}
\begin{align}
	r_i(t) = |\gamma_i| \frac{\lVert L_i \ket{\psi (t)} \rVert^2}{\lVert \ket{\psi(t)} \rVert^2},
	\label{eq:jump_prob}
\end{align} 
for which the quantum jump does not alter the norm $\lVert \ket \psi \rVert\equiv \braket{\psi}^{1/2}$ of the state, i.e.~$\lVert \ket{\psi^{(i)}(t+\delta t)} \rVert = \lVert \ket{\psi(t+\delta t)}\rVert$. Note, however, that for $\gamma_i<0$ this choice implies that the norm increases during the coherent time evolution with $H_\mathrm{eff}$,  $\lVert \ket{\psi^{(0)}(t+\delta t)} \rVert = (1 + \delta t \sum_{\gamma_i<0} r_i(t))  \lVert \ket{\psi(t)} \rVert$ \cite{footnote_norm}. 
This is not the case for the conventional MCWF approach, where $\gamma_i\ge0$.

\textit{Non-Markovian dephasing for single qubit.--} As a proof of principle we implement the PLQT algorithm for a qubit subjected to purely dissipative dynamics,
\begin{figure}[b]
	\includegraphics{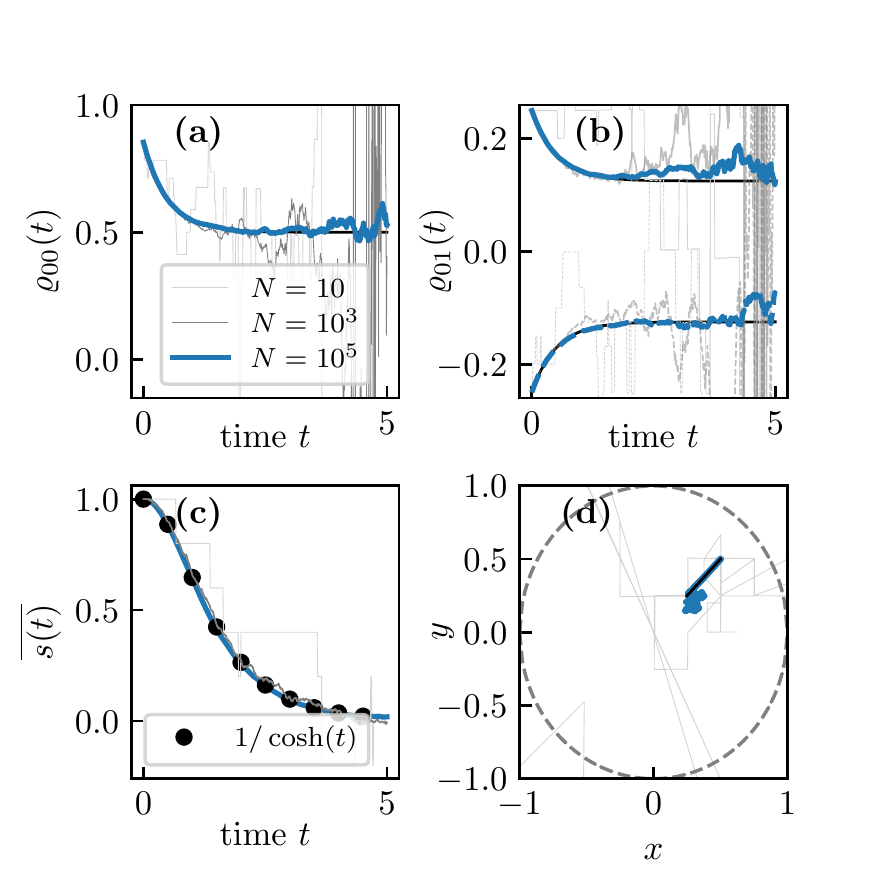}
	\caption{Non-Markovian dynamics [\cref{eq:nonmarkovianqubit}] for density matrix elements $\varrho_{00}$ (a), $\mathrm{Re}\varrho_{01}$ (solid), $\mathrm{Im}\varrho_{01}$ (dashed) (b) (and the Bloch-vector in the $x$-$y$-plane (d). Analytical solution (black) and unraveling with $N=10^5$ PLQTs with time step $ \delta t=0.01$ in blue ($N=10,10^3$ in thin and intermediate grey lines) for an initial Bloch state with $\phi=\Theta=\pi/4$. (c) Shows the averaged sign bit.}
	\label{fig:nonmarkovianqubit}
\end{figure}
\begin{align}
	\begin{aligned}
		\dot{\varrho}(t) &= \frac{1}{2} \Big[ \mathcal{L}_x + \mathcal{L}_y - \tanh(t) \mathcal{L}_z\Big]\varrho(t),
		\label{eq:nonmarkovianqubit}
	\end{aligned}
\end{align} 
with GKSL channels $\mathcal{L}_i\varrho = \sigma_i\varrho\sigma_i- \varrho$, where $\sigma_i$ are Pauli operators, with $\sigma_i^\dagger \sigma_i = \sigma_i^2 = 1$. This equation is known as the eternal non-Markovian master equation \cite{MJWHallJDCresserEAndersson2014,DCruscinskiJPiiloWStrunz2017}. The existence of a negative relaxation rate makes it inaccessible to the standard MCWF, while also the NMQJ approach fails, since $-\tanh(t)<0$ for all~$t$. 

However, for this model the PLQT approach is easily implemented and, because the jump operators are unitary the jump rates are state independent, i.e.~$\lVert \sigma_i \ket{\psi(t)} \rVert^2 = \lVert \ket{\psi(t)} \rVert^2$ leads to a simplification in \cref{eq:jump_prob}, and one has $r_x=r_y=1/2$, $r_z(t)=\tanh(t)/2$. Also the effective Hamiltonian $H_\mathrm{eff} = - \frac{\mathrm i}{2} (1 - \tanh(t)/2)$ entering \cref{eq:stQTdet} is not state-dependent (since it is proportional to the identity).

On average the sign follows the rate equation $\dot{\overline{s(t)}} = -2r_z(t) \overline{s(t)}$, which is solved by $\overline{s(t)} = 1/\cosh(t)$, as shown in \cref{fig:nonmarkovianqubit} (c) \cite{supp_TBeckerCNetzerAEckardt2023}. Quantum jumps are realized in the Bloch-vector representation by reflections at the $y$-$z$-plane and $x$-$z$-plane for $\sigma_x$ and $\sigma_y$, respectively [Fig.~\ref{fig:nonmarkovianqubit} (d)]. The $\sigma_z$ quantum jump is a reflection at the $x$-$y$-plane and, due to the negative relaxation strength, the sign flip is accounted by an additional point reflection at the origin. 

By simulating $N=10^5$ trajectories in \cref{fig:nonmarkovianqubit} (a), (b), we obtain accurate results for the transient dynamics until, at $t_\mathrm{R}\sim 2$, the system reaches the steady state regime. Besides this physical relaxation time, we also find an \emph{algorithmic relaxation time} $t_A \sim 4$, at which the number of negative and positive trajectories become equal and the averaged sign decays to zero [\cref{fig:nonmarkovianqubit} (c)]. Beyond this algorithmic relaxation time, fluctuations are typically increased [\cref{fig:nonmarkovianqubit} (a) and (b)]. This effect can be understood by noting that a stochastic process of a real variable $x_n$ with positive mean $\overline{x}$ will have  bounded fluctuations $\Delta x = \overline{(x-\overline{x})^2}^{-1/2}\leq \overline{x}$ as long as $x_n>0$, whereas $\Delta x$ is not bounded, when $x_n$ can also take negative values. Thus, ideally, $t_A$ should be large compared to the time span of the simulation (which is $t_R$, if we are interested in computing the steady state). The algorithmic relaxation time is determined by the inverse sign-flip rate $r_\text{sf}=\sum_{i,\gamma_i<0} r_i$, e.g.~$t_A =r_\text{neg}^{-1}$ for time-independent $r_\text{sf}$. Thus, we can increase $t_A$ simply by lowering the rates for negative processes with “rates” $\gamma_i<0$ relative to positive ones with $\gamma_i>0$. However, this will also increase the number of trajectories needed for properly sampling those negative-“rate” processes. Thus, before doing this, one should first attempt to rewrite the master equation, so that the relative weight of negative processes is reduced. This can be done for pseudo-Lindblad equations derived from the Redfield equation \cite{TBecker2021}, as we will recapitulate now.

\textit{Redfield dynamics.--} For a microscopic model a master equation is often derived within the Born-Markov-Redfield formalism \cite{Strunz20,Archak22}. We consider a system-bath Hamiltonian of the form $H_\mathrm{tot} = H + \sum_i S_i \otimes B_i + H_i$ with system Hamiltonian $H$ that couples to individual baths $H_i$ where $S_i$ and $B_i$ denote the system and bath coupling operators, respectively. The Redfield equation can then be written in pseudo-Lindblad form \cite{TBecker2021}
\begin{align}
	\begin{aligned}
		\dot \varrho &= - \i[H + H^{\mathrm{LS}}, \varrho] \\ &+  \sum\limits_{i,\sigma=\pm} \sigma \left( L_{i\sigma} \varrho L_{i\sigma}^\dagger - \frac{1}{2} \{ L_{i\sigma}^\dagger L_{i\sigma},\varrho \} \right) ,
	\end{aligned}
\end{align}
with Lamb-shift Hamiltonian $H^\mathrm{LS}= (1/2\i) \sum_i S_i\mathbb{S}_i + \text{H.c.}$, convolution operators $\mathbb{S}_i=\int_0^\infty  d \tau \expval{B(\tau)B} e^{\mathrm{i}H \tau} S_i e^{-\i H \tau}$ and Lindblad-like jump operators
\begin{align}
	L_{i\sigma} = \frac{1}{\sqrt 2} \Bigg[ \lambda_{i}(t) S_i + \sigma \frac{1}{\lambda_{i}(t)} \mathbb{S}_i \Bigg]
\end{align} 
with arbitrary, time-dependent real parameters $\lambda_{i}(t)$.
We see that due to the negative relaxation rates with $\sigma=-1$, the Redfield equation is generally not of GKSL form, unless further approximations are employed in the limit of ultra weak coupling \cite{breuerpetruccione,Archak22,Strunz20} or for high bath temperatures \cite{HWichterich2007,TBecker2021}. 
For a purely Ohmic bath, the choice~\cite{TBecker2021}
\begin{align}
	\lambda_{i, \mathrm{glob}}(t)^2 = \sqrt \frac{\tr\ \mathbb{S}_i^\dagger \mathbb{S}_i}{\tr\ S_i S_i}
\end{align}
minimizes the norm of the negative channels in the pseudo-Lindblad equation globally, i.e.~on average for all states. A further reduction of negative processes can be achieved by state-dependent optimization. Namely, according to \cref{eq:jump_prob}, where (assuming, without loss of generality a normalized state) the rates for negative quantum jumps with $L_{i-}$ are given by $	r_{i-}(t;\lambda_i(t)) = \frac{1}{2} \Big( \lambda_i(t)^2 \lVert S_i \ket{\psi(t)}\rVert^2 + \frac{1}{\lambda_i(t)^2} \lVert \mathbb{S}_i \ket{\psi(t)}\rVert^2 - 2\mathrm{Re} \expval{S_i \mathbb{S}_i}{\psi(t)}  \Big)$.
Thus, the choice,
\begin{align}
	\begin{aligned}
		\lambda_{i, \mathrm{loc}}(t)^2 =  \frac{ \lVert \mathbb{S}_i \ket{\psi(t)} \rVert}{\lVert S_i\ket{\psi(t)} \rVert},
		\label{eq:locopt}
	\end{aligned}
\end{align}
minimizes the rates for negative quantum jumps in the unraveling of the Redfield equation. Since the states in the numerator and the denominator of \cref{eq:locopt} have to be evaluated for evolving the state anyway, this local optimization (which is not described in Ref.~\cite{TBecker2021}) can be implemented efficiently. \\
\begin{figure}[b]
	\centering
	\includegraphics{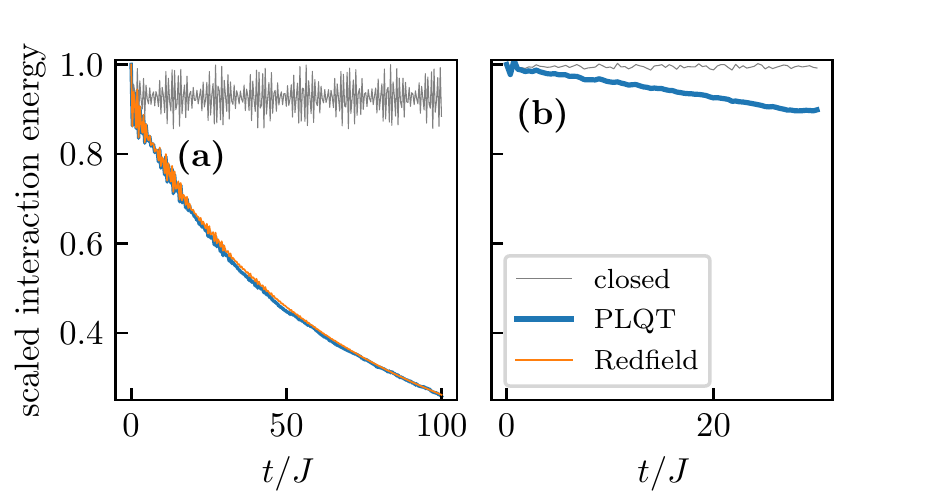}
	\caption{Dynamics of scaled interaction energy of extended Hubbard chain of 2 [8] particles on 4 [13] sites (a) [(b)] with $V/J=7$. We compare the dynamics of the isolated (grey line) and open system with $\gamma/J=0.02$ and $T/J =1$ (blue line for PLQT, thin orange for Redfield equation). The decrease of interactions is related to bath-induced doublon-breaking processes. }
	\label{fig:decaydublons}
\end{figure}
We test our method, using state-dependent minimization \cref{eq:locopt}, for the extended Hubbard chain of spinless fermions,
\begin{align}
	H =  -J\sum\limits_{\ell=0}^{M-1}\Big( a_\ell^\dagger a_{\ell+1} + a_{\ell+1}^\dagger a_\ell \Big) + V \sum\limits_{\ell=0}^{M-1}  a_\ell^\dagger a_\ell a_{\ell+1}^\dagger a_{\ell+1},
\end{align}
with fermionic operators $a_\ell$, tunneling strength $J$ and nearest-neighbour interaction strength $V$. For the dissipator, we have
\begin{align}
	\matrixel{n}{\mathbb{S}_\ell}{m} = \frac{J(\Delta_{nm})}{e^{\Delta_{nm}/T} - 1} \matrixel{n}{S_\ell}{m},
\end{align}
with system operator $S_\ell = a_\ell^\dagger a_\ell$, level splitting $\Delta_{nm} = E_n - E_m$ and bath temperature $T$. We consider a purely Ohmic bath, with spectral density $J(E) = \gamma E$ and coupling strength $\gamma$.

In \cref{fig:decaydublons} we depict the decay of the interaction energy for an initial state in which pairs of adjacent sites are occupied $\ket{\psi(0)} = \ket{011011\dots}$. 
Quench dynamics for such a charge density wave in a spin polarized Fermi Hubbard model have been recently observed experimentally in the Bakr group \cite{WBakr2021}. 
We assume strong interactions, $V/J=7$, for which the dublon pairs can only be broken when the system exchanges energy with the bath. This leads to a decay of the energy of the open system as depicted in \cref{fig:decaydublons} (a), where the transient oscillations are well reproduced.

\textit{Numerical implementation.--} Let us now discuss the numerical implementation of the PLQT approach. Since the trajectories are independent, we run them in parallel. Depending on the physical quantity of interest, let's say observable $A$, it is often reasonable not to store the actual time dependent state as large vectors with complex entries, but rather expectation values $\expval{A}{\psi_n(t)}$ together with the norm $\lVert \psi_n(t) \rVert$ and the sign $s_n(t)$.
While the storage of the trajectory data boils down to a few real numbers, the time evolution requires the full state vector. 
The memory needed for the time integration of a quantum trajectory would grow linear with the state-space dimension $D$, if not only the Hamiltonian, but also the jump operators were sparse. The latter is the case, however, mainly in phenomenological master equations with local jump operators and not for the Redfield equation, so that the memory needed usually scales like $D^2$. The memory needed for integrating the Redfield equation scales equally like $D^2$ (since it is sufficient to store and apply the jump operators rather than the full superoperator). Nevertheless, we find that the memory requirement for quantum trajectories to be much lower than that for integrating the master equation. In \cref{fig:speedup_memory} the required memory, (a) and (b), and single-time-step run time, (c) and (d), is compared for solving the full Redfield master equation (blue) and a single trajectory (red).
\begin{figure}[t]
	\centering
	\includegraphics{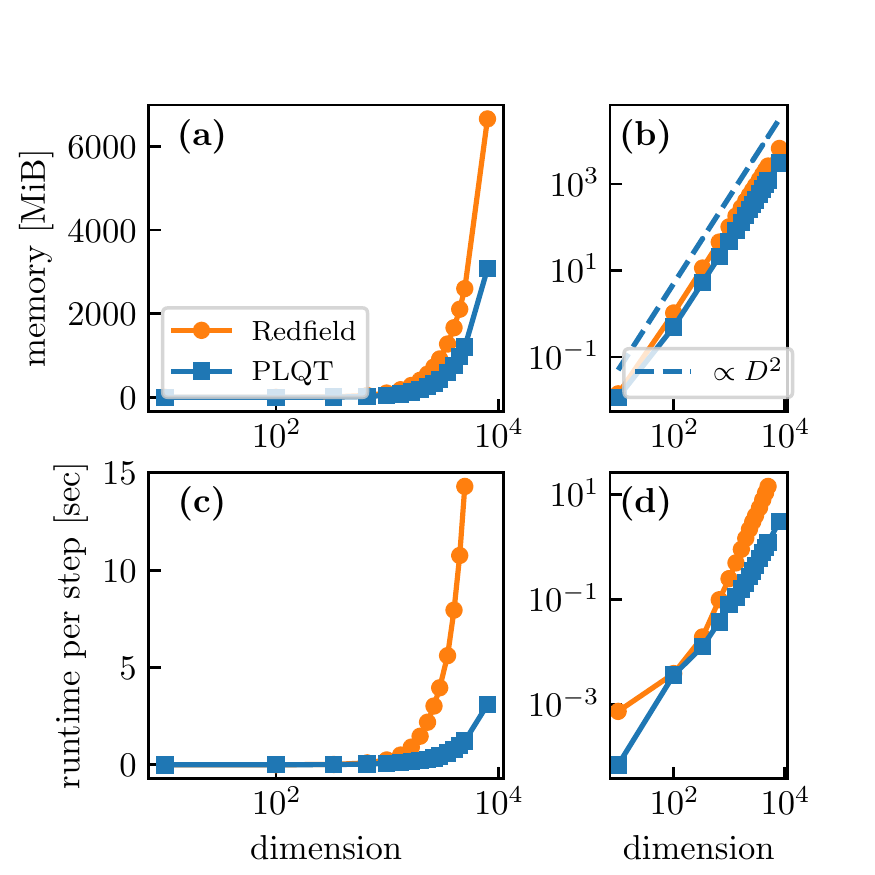}
	\caption{Required memory and single-time-step run time for both a single trajectory (blue) and the Redfield master equation (red). The data was obtained for an Intel Core i9-10900 processor with up to $5.2\,\mathrm{GHz}$.}
	\label{fig:speedup_memory}
\end{figure}
We find that the required memory is noticeably reduced for the quantum trajectory simulation, even though, as discussed above, it still scales like $D^2$. (The latter is not specific to our approach, but generically the case also for other forms of quantum trajectory simulations). For the run time the relative reduction is even stronger and shows different scaling with $D$. Essentially the difference are two matrix-matrix products needed for the Redfield integration and one matrix-vector product for the PLQTs. 
Note that the unravelling can also be combined with matrix product states (e.g.~\cite{Verstraete_2008,USchollwoeck2011}). It is interesting to see in how far such an approach would compare to a representation of the density operator by matrix product operators (e.g. Refs.~\cite{XXuJThinga2019,Weimer2021}).

Finally, we would like to mention that while finalizing this manuscript, a new approach for microscopically deriving accurate local quantum master equations was proposed \cite{schnellGlobalBecomesLocal2023}. It is a very promising perspective to apply the the PLQT to this equation, which can be broad into pseudo-Lindblad form \cite{schnellGlobalBecomesLocal2023}. Namely, it will give rise to equations of motion involving spars matrices for Hamiltonian and jump operators only. As a result, the overall memory requirement will scale linearly with the Hilbert-space dimension $D$ only. Moreover, this could allow for efficient matrix-product state approaches for individual trajectories. 

\textit{Conclusion.--} We have developed an efficient unraveling of master equations of pseudo-Lindblad form, which includes the Redfield equation as an important case \cite{TBecker2021}. Different from previous approaches, it requires a minimal extension of the state space by one classical sign bit only, is applicable also for dissipation strengths that are always negative, does not require matrix diagonalization during the time integration, and allows for a parallel implementation, since all trajectories are independent from each other. We believe that it will be a useful tool for the simulation of open many-body systems beyond ultra weak system-bath coupling. In future work it will be interesting to systematically investigate the impact of negative dissipation strengths $\gamma_i$ on the required ensemble size and the optimal choice of the corresponding rates for efficient simulation. Moreover, our algorithm should be compared to the influence-martingale approach \cite{DonvilMGinanneschi2022}. Finally, the combination of PLQT simulations with the recently derived local quantum master equation \cite{schnellGlobalBecomesLocal2023} opens a promising route for the efficient simulation of open many-body quantum systems.

\begin{acknowledgments}
	We thank Francesco Intravaia, Charlotte Maurer and Alexander Schnell for fruitful discussions. 
	This research was funded by by the Deutsche Forschungsgemeinschaft (DFG) via the Research Unit FOR 2414 under project No. 277974659.
\end{acknowledgments}

\bibliography{refsgntraj}

\begin{thebibliography}{51}
\bibitem [{foo()}]{footnote_det}%
  \BibitemOpen
  \href@noop {} {}\bibinfo {note} {By Taylor expanding the denominator in
  Eq.~[4] to first order in the small time step $\delta t$, $1/\sqrt{1- \delta
  t \sum_i r_i(t)} \simeq \delta t/2 \sum_i r_i(t)$ the deterministic evolution
  can equivalently be formulated as first order differential equation
  $\partial_t \ket*{\psi^{(0)}(t)} =\left( -\mathrm{i} H_\mathrm{eff}(t) + 1/2
  \sum_i r_i(t) \right) \ket*{\psi^{(0)}(t)} $.}\BibitemShut {Stop}%
\end{thebibliography}

\clearpage

\title{Supplemental Material for\\``Quantum trajectories for time-local non-Lindblad master equations"}
\maketitle

\setcounter{equation}{0}
\makeatletter
\renewcommand{\theequation}{A\arabic{equation}}

\onecolumngrid

\section{Pseudo-Lindblad quantum trajectory unraveling in Ito-formalism}
The starting point is the pseudo-Lindblad master equation
\begin{align}
	\dot \varrho = -i[H, \varrho] + \sum\limits_{i} \gamma_i \left( L_i \varrho L_i^\dagger - \frac{1}{2} \{ L_i^\dagger L_i,\varrho \} \right) ,
\end{align}
with jump operators $L_i$ and relaxation strength $\gamma_i$, which we explicitly allow to be negative. For convenience, in the following we absorb the magnitude of relaxation strength into the jump operator $L_i \propto \sqrt{|\gamma_i|}$ and denote the sign by $\sigma_i=\gamma_i/|\gamma_i|=\pm 1$,
\begin{align}
	\dot \varrho = -i[H, \varrho] + \sum\limits_{i} \sigma_i \left( L_i \varrho L_i^\dagger - \frac{1}{2} \{ L_i^\dagger L_i,\varrho \} \right) .
	\label{eq:app_pseudo_lindblad}
\end{align}
Each channel $i$ gives rise to a stochastic process, which we can unravel with pseudo-Lindblad quantum trajectories (PLQT). In the Ito-formalism the change of the trajectory is described as follows:
\begin{align}
	\ket{ \mathrm d\psi} = \sum_i \mathrm d N_i \Big(\frac{\sigma_i L_i }{\sqrt{ r_i(t)}} \ket{\psi} - \ket \psi \Big) + \mathrm d t \Big( \sum_i \frac{ r_i(t)}{2} - \i H_\mathrm{eff} \Big) \ket \psi ,
	\label{eq:app_ito_formalism}
\end{align}
with $\mathrm d N_i$ describing independent jump processes. For quantum jumps $\mathrm d N_i$ take the values $1$ or $0$, depending on whether a jump takes place or not. The average value
\begin{align}
	\overline{\mathrm d N_i} =  r_i(t) \mathrm d t
\end{align}
defines the jump rate $r_i(t)>0$, which is a free parameter because it later cancels out in the ensemble average. The second term in \cref{eq:app_ito_formalism} describes deterministic evolution, 
\begin{align}
	\partial_t \ket{\psi^{(1)}} = \Big(\sum_i \frac{ r_i(t)}{2} - \i H_\mathrm{eff} \Big) \ket{\psi^{(1)}}  ,
	\label{eq:app_deterministic}
\end{align}
where $H_\mathrm{eff} = H - \frac{\i}{2} \sum_i \sigma_i L_i^\dagger L_i$ with anti Hermitian contribution is often regarded as effective Hamiltonian. 

For the deterministic time evolution the sign is directly encoded in the effective Hamiltonian, whereas for the quantum jump one cannot differentiate between positive and negative relaxation strengths. This can be seen from the fact that $-L_i\ket\psi \simeq L_i\ket\psi$ differ by a global phase factor only and, thus, are equivalent. In other words for a pure state $\ketbra{\psi}$ the relative sign in the jumps $(-L_i)\ketbra{\psi}(-L_i)^\dagger$ would simply cancel. In order to recover the relative sign the idea is to create an asymmetric ensemble $\dyad*{\psi}{\tilde\psi}=s \dyad{\psi}{\psi}$ where $\ket{\tilde \psi}$ is a copy of the trajectory $\ket \psi$, except that the sign after a quantum jump is not changed,
\begin{align}
	\ket{ \mathrm d\tilde \psi} = \sum_i \mathrm d N_i \Big(\frac{L_i }{\sqrt{ r_i(t)}} \ket{\tilde \psi} - \ket{\tilde \psi} \Big) + \mathrm d t \Big( \sum_i \frac{ r_i(t)}{2} - \i H_\mathrm{eff} \Big) \ket{\tilde \psi}.
	\label{eq:app_ito_formalism_copy}
\end{align}
This is the same asymmetric unraveling proposed by Breuer and coworkers in reference \cite{breuerpetruccione1999} for the special case that the trajectories only differ by a relative sign. That is why it is not necessary to simulate $\ket{\tilde \psi}$ but that it is sufficient to store the information about the additional sign bit. 

The rest of the proof follows the standard arguments of any quantum jump unraveling. A nice overview is given by I.\ Kondov and coworkers in reference \cite{IKondovMschreiber2003}. To show that the unraveling yields on average the pseudo-Lindblad equation (\ref{eq:app_pseudo_lindblad}), we calculate an infinitesimal change of one signed trajectory,
\begin{align}
	\mathrm d (s \dyad{\psi}) &= \mathrm d \dyad*{\psi}{\tilde\psi} = \dyad*{\mathrm d \psi}{\tilde\psi} + \dyad*{\psi}{\mathrm d \tilde \psi} + \dyad*{\mathrm d \psi}{\mathrm d \tilde \psi}.
\end{align}
Note that, for stochastic processes the quadratic term also contributes to lowest order. The first two terms give
\begin{align}
	\begin{split}
		\dyad*{\mathrm d \psi}{\tilde\psi} + \dyad*{\psi}{\mathrm d \tilde \psi} &= \sum_i \mathrm dN_i \Big(\frac{\sigma_i L_i \dyad*{ \psi}{\tilde\psi} }{\sqrt{ r_i(t)}} + \frac{\dyad*{ \psi}{\tilde\psi} L_i^\dagger  }{\sqrt{ r_i(t)}} - \dyad*{ \psi}{\tilde\psi} \Big) \\ & - \sum_i \mathrm dN_i \dyad*{ \psi}{\tilde\psi}+ \mathrm dt \Bigg(\sum_i  r_i(t) \dyad*{ \psi}{\tilde\psi} - \i H_\mathrm{eff} \dyad*{ \psi}{\tilde\psi} + \dyad*{ \psi}{\tilde\psi}\i H_\mathrm{eff}^\dagger \Bigg) ,
	\end{split}
	\label{eq:app_hermitianterms}
\end{align}
and essentially contribute to the effective Hamiltonian. Note that on average the first terms in the second line cancels, since $\overline{\mathrm d N_i} = r_i\mathrm d t$.

For the quadratic term $\dyad*{\mathrm d \psi}{\mathrm d \tilde \psi}$ only stochastic processes $\mathrm d N_i$ contribute. Since the processes are independent and for single realizations one has $\mathrm dN_i=0,1$, it follows that $\mathrm dN_i \mathrm dN_j = \delta_{ij}\mathrm dN_i$. We then get
\begin{align}
	\begin{split}
		\dyad*{\mathrm d \psi}{\mathrm d \tilde \psi} & = \sum_i \mathrm dN_i^2 \Big(\frac{\sigma_iL_i \ket{\psi} }{\sqrt{ r_i(t)}} - \ket{\psi} \Big)\Big(\frac{ \bra{\tilde \psi}L_i^\dagger }{\sqrt{ r_i(t)}} - \bra{\tilde \psi} \Big)  \\&= \sum_i \mathrm dN_i\ \sigma_i \frac{L_i \dyad*{ \psi}{\tilde\psi} L_i^\dagger}{ r_i(t)} - \sum_i \mathrm dN_i \Big(\frac{\sigma_i L_i \dyad*{ \psi}{\tilde\psi}}{\sqrt{ r_i(t)}} + \frac{\dyad*{ \psi}{\tilde\psi} L_i^\dagger  }{\sqrt{ r_i(t)}} - \dyad*{ \psi}{\tilde\psi} \Big) ,
	\end{split}
\end{align}
where the latter terms cancel the first line in \cref{eq:app_hermitianterms}, and we arrive at
\begin{align}
	\mathrm d (\dyad*{ \psi}{ \tilde\psi}) =\mathrm dt\, \Big[- \i H_\mathrm{eff} \dyad*{ \psi}{ \tilde\psi} + \dyad*{ \psi}{ \tilde\psi} \i H_\mathrm{eff}^\dagger\Big] +  \sum_i  \frac{\mathrm dN_i }{ r_i(t) } \sigma_i L_i\dyad*{ \psi}{ \tilde\psi} L_i^\dagger .
\end{align}
By taking the ensemble average we replace $\overline { \mathrm d N_i }= r_i(t) \mathrm d t$ and arrive at the pseudo-Lindblad equation (\ref{eq:app_pseudo_lindblad}) for the density matrix $\varrho \equiv \overline{\dyad*{ \psi}{ \tilde\psi}}$.

\section{Norm conservation and choice of jump rates}
As shown above the jump rates $r_i(t)$ are free parameters of the unraveling because they cancel out in the ensemble average \cite{IKondovMschreiber2003}.  This is similar to the relaxation strengths in the master equation \cref{eq:app_pseudo_lindblad}, which can always be redefined by rescaling the jump operators, as we employed it above to bring the master equation to the form \cref{eq:app_pseudo_lindblad}. However, the jump rates influence the dynamics on the level of single trajectories and typically one chooses state-dependent jump rates (see main text)
\begin{align}
	r_i(t) = \frac{\lVert L_i\psi (t) \rVert^2}{\lVert\psi(t)\rVert^2},
\end{align} 
which yield quantum jumps that do not change the norm of the state:
\begin{align}
	\ket{\psi(t) } \rightarrow \frac{L_i \ket{\psi(t)}}{\sqrt{r_i(t)}} = \frac{L_i \ket{\psi(t)}}{\lVert L_i\psi (t) \rVert} \lVert \psi (t) \rVert.
\end{align} 
Also for state-dependent jump rates the deterministic time evolution \cref{eq:app_deterministic} is non-linear, which for the MCWF compensates for the loss of norm under the effective Hamiltonian. In turn, for master equations with negative signs the norm increases according to
\begin{align}
	\begin{split}
		\partial_t \lVert \psi^{(1)} \rVert^2 &= \ip*{\psi^{(1)}}{\partial_t \psi^{(1)}} + \ip*{\partial_t \psi^{(1)}}{\psi^{(1)}} \\ &= \underbrace{\matrixel*{\psi^{(1)}}{-\i H}{\psi^{(1)}} + \matrixel*{\psi^{(1)}}{\i H}{\psi^{(1)}}}_{=0}+ \sum_{i} \matrixel*{\psi^{(1)}}{( r_i(t) - \sigma_i L_i^\dagger L_i)}{\psi^{(1)}} \\
		&= 2 \sum_{i, \sigma_i=-1} \matrixel*{\psi^{(1)}}{L_i^\dagger L_i }{\psi^{(1)}},
	\end{split}
	\label{eq:app_deterministic_norm}
\end{align}
where the sum is taken over all channels with negative relaxation rate. 
Alternatively, one may choose different jump rates $r_i(t)>0$ such that the deterministic time evolution is norm preserving. The condition follows from \cref{eq:app_deterministic_norm} to be
\begin{align} 
	\sum_i r_i(t) = \sum_i \sigma_i \frac{\lVert L_i \psi \rVert^2 }{ \lVert \psi \rVert^2}. 
\end{align}	
For unitary jump operators the rates can be chosen state-independent. 

\begin{figure}[t]
	\includegraphics{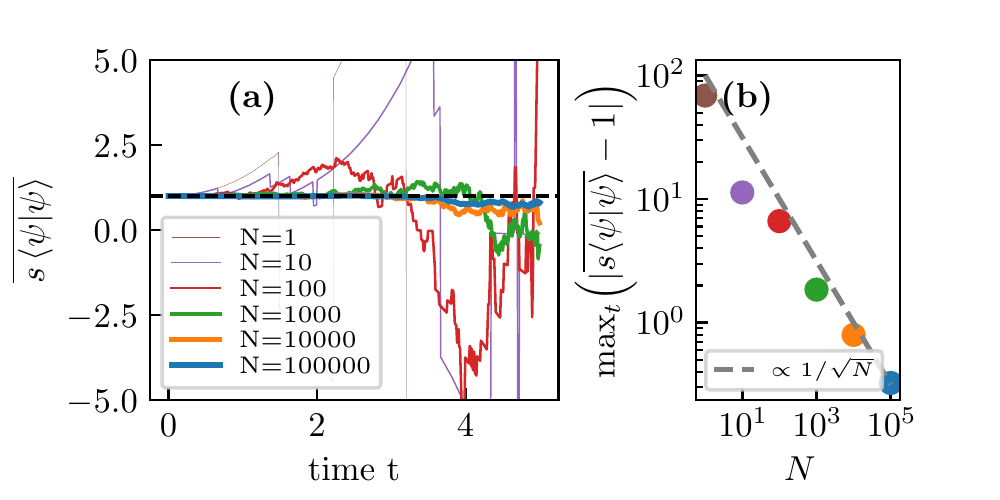}
	\caption{We simulate PLQTs for the eternal non-Markovian master equation discussed in the main text. (a) Dynamics of ensemble average $\overline{s\, \langle \psi|\psi \rangle}$ for increasing number $N$ of trajectories (thicker lines). (b) Maximal deviation of the average from the value $1$ during the time evolution as a function of ensemble size $N$.}
	\label{fig:convergence_trace}
\end{figure}

For the eternal non-Markovian master equation discussed in the main text the condition reads $\sum_i r_i(t) = 1 - \tanh(t)/2$, so that one possible choice is
\begin{align}
	r_x = r_y = \frac{1}{4}, \qquad r_z = \frac{1 - \tanh(t)}{2},
\end{align}
as compared to the choice in the main text [$r_x=r_y=1/2$, $r_z=\tanh(t)/2$]. Again, the rates for $\sigma_x$ and $\sigma_y$-jumps are taken to be equal. The $\sigma_z$ jump, which corresponds to the negative relaxation strength $\gamma_i<0$, happens more often in the beginning but vanishes in the static limit. Note, if the deterministic time evolution is norm conserving, the quantum jumps are not. For the qubit this would give quantum jumps 
$\ket{\psi} \rightarrow 2 \sigma_x \ket{\psi}$, $\ket{\psi} \rightarrow 2 \sigma_y \ket{\psi}$ and $\ket{\psi} \rightarrow \sqrt 2 \sigma_z \ket{\psi}/\sqrt{1-\tanh(t)}$. One concludes that for negative relaxation strengths, there is no choice for jump rates $r_i(t)>0$ for which the trajectories remain normalized.
The increase of the norm, however, is cancelled by trajectories with opposing sign, such that on average the sign is preserved. This is shown in \cref{fig:convergence_trace} (a), where we plot the dynamics of the ensemble average $\frac{1}{N} \sum_n^N s_n \langle \psi_n|\psi_n \rangle$ for increasing number $N$ of trajectories. For proving this statement, note that on average the ensemble of trajectories evolves according to the master equation, which is trace-preserving, i.e.~$\tr \overline{s\, \dyad \psi} =  \overline{s\, \langle \psi|\psi \rangle} = 1$. For finite $N$ the fluctuations are on the order of $1/\sqrt{N}$ [\cref{fig:convergence_trace} (b)]. They can be removed by explicitly normalizing the ensemble as mentioned in the main text.

\section{Dynamics of the averaged sign bit}
Following the arguments in the main text, a trajectory $\{ \ket{ \psi(t)}, s(t) \}$ probabilistically changes its sign with the rate 
\begin{align}
	R(t) \equiv \sum_{i, \gamma_i < 0} r_i(t),
\end{align}
where the sum runs over all jump rates $r_i(t)$ that correspond to a negative $\gamma_i<0$. The dynamics of the sign then follows the update scheme 
\begin{align}
	s(t + \delta t) = \begin{cases}
		-s(t) & \text{with probability } R(t) \delta t \\
		s(t) & \text{with probability } 1 - R(t) \delta t. 
	\end{cases}
\end{align}
For the averaged sign $\overline s(t)$ one takes the weighted average
\begin{align}
	\overline s(t+\delta t) = - R(t)\delta t\, \overline s(t) + \left( 1-R(t) \delta t \right) \, \overline s(t) 
\end{align}
from which we can deduce the first order differential equation
\begin{align}
	\dot{\overline{s}}(t) = - 2 R(t)\ \overline{s}(t),
\end{align}
where we have used the limit $\dot{\overline{s}}(t)=\lim\limits_{\delta t\to 0} (s(t+\delta t) - s(t))/\delta t$.

Provided $R(t)$ is a known, smooth and state-independent function, the averaged sign follows an exponential decay 
\begin{align}
	\overline{s}(t) = \exp\Big[-2 \int_0^t R(\tau) \mathrm d \tau \Big].
\end{align}
This is the case, e.g., for the eternal master equation discussed in the main text. However, even if the decay rate depends on the particular state, qualitatively the net sign decays exponentially on a time scale that is given by the negative relaxation rate. Therefore, eventually for sufficiently long times the number of positive and negative trajectories will be balanced (up to statistical fluctuations). As a result $\bar{s}$ vanishes for times that are long compared to an \emph{algorithmic relaxation time}.

\begin{figure}[t]
	\begin{tikzpicture} [>=latex]
		\node [draw,rectangle] (Start) {$p_+$};
		\node [draw,rectangle] (Finish) [right=2cm] {$p_-$};
		\draw [->,out=90,in=90,looseness=1] (Start.north) to node[below]{rate $R$}  (Finish.north);
		\draw [->,out=-90,in=-90,looseness=1] (Finish.south) to node[above]{rate $R$} (Start.south);
	\end{tikzpicture}
	\caption{Let $R$ denote the rate under which trajectories change their sign. 
		For long times, the algorithm will always equilibrate to a state having (up to statistical fluctuations) the same number of trajectories with positive and negative sign. This can be readily seen from the simple rate equation, $\dot p_+ = -R\, (p_+ - p_-)$ and $\dot p_-=-\dot p_+$, for the relative amount of positive and negative trajectories, respectively.}
	\label{fig:app_signrateeq}
\end{figure}
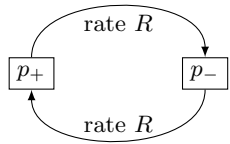


\end{document}